\documentclass[epj,referee]{svjour}
\usepackage{graphics}
\begin{document}
\title{The spontaneous generation of magnetic and chromomagnetic fields
  at high temperature in the Standard Model}
\authorrunning{V.I.Demchik and V.V.Skalozub}
\titlerunning{The spontaneous generation of
magnetic and chromomagnetic fields in the SM}
\author{V.I. Demchik\inst{1}\thanks{\emph{E-mail:} vadimdi@yahoo.com}
\and V.V. Skalozub\inst{2}\thanks{\emph{E-mail:} skalozub@ff.dsu.dp.ua}
}                     
\offprints{}          
\institute{Dniepropetrovsk National University, Dniepropetrovsk, Ukraine
\and Dniepropetrovsk National University, Dniepropetrovsk, Ukraine}

\date{Received: 22.10.2001 / Revised version: 22.10.2001}
\abstract{ The spontaneous generation of the magnetic and
chromomagnetic fields at high temperature is investigated in the
Standard Model. The consistent effective potential including the
one-loop and the daisy diagrams of all boson and fermion fields is
calculated. The mixing of the generated fields due to the quark
loop diagram is studied in detail. It is found that the quark
contribution increases the magnetic and chromomagnetic field
strengths as compared with the separate generation of fields. The
magnetized vacuum state is stable due to the  magnetic gauge field
masses included in the daisy diagrams. Some applications of the
results obtained are discussed.
}
\maketitle

\section{Introduction}
One of interesting problems of nowadays high energy phy-
sics is generation of strong magnetic fields in the early universe.
Different mechanisms of producing the fields at different stages
of the universe evolution have been proposed (see, for instance,
the surveys \cite{Ken,Ken2,Dar}) and the influence of fields on
various processes was discussed. In particular, the primordial
magnetic fields, being implemented in cosmic plasma, may serve
as the seed source of the present extra galaxy fields.

 One of the mechanisms is a spontaneous vacuum magnetization at high
temperature. It was  investigated already for the case of pure $SU(2)$
gluodynamics in Refs. \cite{Enq,Sta,Once}, where it
has been shown the possibility of this phenomenon. The stability
of the magnetized vacuum was also studied \cite{Once}. As it is
well known, the magnetization takes place for the non-abelian
gauge fields due to a vacuum dynamics \cite{Sv}. In fact, this is
one of the distinguishable features of asymptotically free
theories. In the papers mentioned the fermions were not taken into
consideration. However, they may affect the vacuum state due to
loop corrections in strong magnetic fields at high temperature.

 In the present paper the spontaneous vacuum magnetization
is investigated in the standard model (SM) of elementary
particles. All boson and fermion fields are taken into
consideration. In the SM there are two kind of non-abelian gauge
fields - the $SU(2)$ weak isospin gauge fields responsible for
weak interactions and the $SU(3)$ gluons mediating strong
interactions. The quarks possess both the electric and colour
charges. So, they have to mix the chromomagnetic and the ordinary
magnetic fields due to vacuum loops. Because of this mixing some
specific configurations of the fields must be produced at high
temperature. To elaborate this picture quantitatively, we
calculate the effective potential (EP) including the one-loop and
the daisy diagram contributions in the constant abelian
chromomagnetic and magnetic fields, $H_c=const$ and $H=const$,
at high temperatures. This type of resummations guarantees a
vacuum stability if one takes into consideration the one-loop
temperature masses of the transversal gauge fields in the external
fields considered \cite{Once}. So, we will used this approximation
in what follows. Since an abelian magnetic hypercharge field is
not generated spontaneously, in what follows we shall consider the
non-abelian component of the magnetic field. The mechanisms of
hypermagnetic field generation has been discussed in Refs.
\cite{Gio,Jo}. It will be shown that at high temperatures either
the strong  magnetic or the chromomagnetic fields are generated.
They are stable in the approximation adopted due to the magnetic
masses of $m^2_{transversal} \sim (gH)^{1/2} T$ of the gauge field
transversal modes \cite{SkSt}. In this way the consistent picture
of the magnetized vacuum state in the SM at high temperature can
be derived.

The paper content is as follows. In sect. 2 the contributions of
bosons and fermions to the EP $v^\prime(H,T)$ of external magnetic
and chromomagnetic fields are calculated in the form convenient
for numeric investigations. In sect. 3 the field strengths are
calculated. Discussion and concluding remarks are given in sect. 4.

\section{Basic Formulae}
The SM Lagrangian of the gauge boson sector is (see, for example,
\cite{Cheng})
\begin{eqnarray}
L=-\frac14F^\alpha_{\mu\nu}F^{\mu\nu}_\alpha-\frac14G_{\mu\nu}G^{\mu\nu}-
\frac14{\bf F}^\alpha_{\mu\nu}{\bf F}^{\mu\nu}_\alpha,
\end{eqnarray}
where the standard notations are introduced
\begin{eqnarray}
F^\alpha_{\mu\nu}=\partial_\mu A^\alpha_\nu-\partial_\nu
A^\alpha_\mu+g\epsilon^{abc}A^b_\mu A^c_\nu,\\\nonumber
G_{\mu\nu}=\partial_\mu B_\nu-\partial_\nu B_\mu,\\\nonumber
{\bf F}^\alpha_{\mu\nu}=\partial_\mu {\bf A}^\alpha_\nu-\partial_\nu
{\bf A}^\alpha_\mu+g_sf^{abc}{\bf A}^b_\mu {\bf A}^c_\nu.
\end{eqnarray}
The fields corresponding to the $W-$, $Z-$bosons and photons,
respectively, are
\begin{eqnarray}
W_\mu^\pm=\frac{1}{\sqrt{2}}(A^1_\mu\pm iA^2_\mu),\\\nonumber
Z_\mu=\frac{1}{\sqrt{g^2+g^{\prime 2}}}(gA^3_\mu-g^\prime B_\mu),\\\nonumber
A_\mu=\frac{1}{\sqrt{g^2+g^{\prime 2}}}(g^\prime A^3_\mu+g B_\mu)
\end{eqnarray}
and ${\bf A^\alpha_\mu}$ is the gluon field.

To introduce an interaction with the magnetic and chromomagnetic
fields we replace all derivatives in the Lagrangian by the covariant ones,
\begin{equation}
\label{Cod}
\partial_\mu\to
{\cal D}_\mu=\partial_\mu+ig\frac{\tau^\alpha}{2}A^\alpha_\mu
+ig_s\frac{\lambda^\alpha}{2}{\bf A}_\mu^\alpha.
\end{equation}
Here $\tau^\alpha$ and $\lambda^\alpha$ stand for the Pauli and
the Gell-Mann matrices, respectively.

In the $SU(2)$ sector of the SM there is only one magnetic field -
the third projection of the gauge field. In the $SU(3)_c$ sector
there are two possible chromomagnetic fields connected with the
third and eighth generators of the $SU(3)$.

For simplicity, in what follows we shall consider the field
associated with the third generator of the $SU(3)_c$.

The introduction of an interaction with classical magnetic and
chromomagnetic fields as usually is doing by splitting the
potentials in two parts
\begin{eqnarray}
A_\mu=\bar{A}_\mu+A^R_\mu,\\\nonumber
{\bf A}_\mu=\bar{\bf A}_\mu+{\bf A}^R_\mu,
\end{eqnarray}
where $A^R$ and ${\bf A}^R$ describe the radiation fields and $\bar{A}=$
$(0,0,Hx^1,0)$ and $\bar{\bf A}=(0,0,{\bf H}_3x^1,0)$
correspond to the constant magnetic ~ and ~ chromomagnetic fields
directed along the third axes in the space and in the internal
colour and isospin spaces.

We used the general relativistic renormalizable gauge which
is set by the following gauge-fixing conditions \cite{VVk}
\begin{eqnarray}
\label{gauge1}
\partial_\mu W^{\pm \mu}\pm ie\bar{A}_\mu W^{\pm \mu} \mp
i \frac{g\phi_c}{2\xi}\phi^\pm=C^\pm(x),\\
\nonumber
\partial_\mu Z^\mu-\frac{i}{\xi^\prime}(g^2+g^{\prime
2})^{1/2}\phi_c \phi_Z=C^Z(x),\\
\nonumber
\partial_\mu {\bf A}^\mu + ig_s \bar{\bf A}={\bf C}(x),
\end{eqnarray}
where $e=g~sin \theta_W$, $tan\theta_W=g^\prime/g$, $\phi^\pm$ and
$\phi_Z$ are the Goldstone fields, $\xi$ and $\xi^\prime$ are the
gauge fixing parameters, $C^\pm$ and $C^Z$ are arbitrary functions
and $\phi_c$ is a value of the scalar field condensate. Setting
$\xi,\xi^\prime=0$ we choose the unitary gauge. In the restored phase
the scalar field condensate $\phi_c=0$ and the equations (\ref{gauge1})
are simplified.

The values of the macroscopic magnetic and chromomagnetic fields
generated at high temperature will be calculated by minimization
of the thermodynamics potential.

The thermodynamic potential $\Omega$ of the model is
\begin{eqnarray}
\label{Omeg}
\Omega=-\frac{1}{\beta}log~Z,\\
Z=Tr~exp(-\beta{\cal{H}}),
\end{eqnarray}
where $Z$ is the partition function, ${\cal{H}}$ is the
Hamiltonian of the system. The trace is calculated over all
physical states.

To obtain the EP one has to rewrite (\ref{Omeg}) as a sum in
quantum states calculated near the nontrivial classical solutions
$A^{ext}$ and ${\bf A}^{ext}$. This procedure is well-described
in literature (see, for instance, \cite{Once,Kap,Carr})
and the result can be written in the form:
\begin{eqnarray}
V=V^{(1)}(H,{\bf H}_3,T)+V^{(2)}(H,{\bf H}_3,T)+...\\\nonumber
+V_{daisy}(H,{\bf H}_3,T)+...,
\end{eqnarray}
where $V^{(1)}$ is the one-loop EP, the other terms present the
contributions of two-, three-, etc. loop corrections.

Among these terms there are ones responsible for dominant
contributions of long distances at high temperature - so-called
daisy or ring diagrams (see, for example, \cite{Kap}). This part
of the EP, $V_{daisy}(H,{\bf H}_3,T)$, is nonzero in the case when
massless states appear in a system. The ring diagrams have to be
calculated when the vacuum magnetization at finite temperature is
investigated. Really, one first must assume that the fields are
nonzero, calculate EP $V(H,{\bf H}_3,T)$ and after that check
whether its minimum is located at nonzero $H$ and ${\bf H}_3$. On
the other hand, if one investigates problems in the applied
external fields, the charged fields become massive with the masses
depending on $\sim(gH)^{1/2}$, $\sim(g_s{\bf H}_3)^{1/2}$  and
have to be omitted.

The one-loop contribution into EP is given by the expression
\begin{equation}
\label{Trace}
V^{(1)}=-\frac12 Tr~log~G^{ab},
\end{equation}
where $G^{ab}$ stands for the propagators of all quantum
fields $W^\pm$, ${\bf A}$, $\ldots$ in the background fields $H$
and ${\bf H}_3$. In the proper time formalism, $s$-representation,
the calculation of the trace
can be carried out in accordance with the formula \cite{Schw}
\begin{equation}
Tr~log~G^{ab}=-\int_0^\infty\frac{ds}{s}tr~exp(-isG^{-1}_{ab}).
\end{equation}
Details of calculations based on the $s$-representation and
formula (\ref{Green}) can be found in Refs. \cite{Cab}, \cite{Rez}
and \cite{Ska}.

We make use the method of Ref. \cite{Cab} allowing in a natural
way to incorporate the temperature into this formalism. A basic
formula of Ref. \cite{Cab} connecting the Matsubara Green
functions with the Green functions at zero temperature is needed,
\begin{eqnarray}
\label{Green}
G^{ab}_k(x,x^\prime;&&T)=\\\nonumber
&&\sum^{+\infty}_{-\infty}
(-1)^{(n+[x])\sigma_k}G^{ab}_k(x-[x]\beta u,x^\prime-n\beta u),
\end{eqnarray}
where $G^{ab}_k$ is the corresponding function at $T=0$,
$\beta=1/T$, $u=(0,0,0,1)$, $[x]$ denotes an integer
part of $x_4/\beta$, $\sigma_k=1$ in the case of physical fermions
and $\sigma_k=0$ for boson and ghost fields. The Green functions
in the right-hand side of (\ref{Green}) are the matrix elements of
the operators $G_k$ computed in the states $\vert
x^\prime,a\rangle$ at $T=0$, and in the left-hand side the
operators are averaged over the states with $T\neq 0$. The
corresponding functional spaces $U^0$ and $U^T$ are different but
in the limit of $T\to 0$ $U^T$ transforms into $U^0$.

The terms with $n=0$ in Eqs. (\ref{Green}),
(\ref{Trace}) give the zero temperature expressions for the Green
functions and the effective potential $V^\prime$, respectively.
So, we can split it into two parts:
\begin{equation}
V^\prime(H,{\bf H}_3,T)=V^\prime(H,{\bf H}_3)
+V^\prime_\tau (H,{\bf H}_3,T).
\end{equation}
The standard procedure to account for the daisy diagrams is to
substitute the tree level Matsubara Green functions in
(\ref{Trace}) $[G^{(0)}_i]^{-1}$ by the  full propagator
$G^{-1}_i=[G^{(0)}_i]^{-1}+\Pi(H,T)$ (see for details \cite{Once},
\cite{Kap}, \cite{Carr}), where the last term is polarization
operator at finite temperature in the field taken at zero
longitudinal momentum $k_l=0$.

Passing the detailed calculations we can notice that the exact
one-loop EP will transformed into EP, which contains the daisy
diagrams as well as one-loop diagrams, by adding term contained
the temperature dependent mass of particle to the exponent.

It is convenient for what follows to introduce the dimensionless
quantities: $x=H/H_0$ $(H_0=M_W^2/e)$, $y={\bf H}_3/{\bf H}_3^0$
$({\bf H}_3^0=M_W^2/g_s)$, $B=\beta M_W$, $\tau=1/B=T/M_W$, $v=V/H_0^2$.

The total EP in our consideration consists of the several terms
\begin{eqnarray}
\label{vprime}
v^\prime&=&\frac{x^2}{2}+\frac{y^2}{2}+v^\prime_{leptons}+v^\prime_{quarks}
\\\nonumber&&+v^\prime_{W-bosons}+v^\prime_{gluons}.
\end{eqnarray}

These terms can be exactly written for SM fields
(in dimensionless variables):

- leptons
\begin{eqnarray}
v^\prime_{leptons}=-\frac{1}{4 \pi^2}\sum_{n=1}^{\infty}
(-1)^n \int_0^\infty
\frac{ds}{s^3}
\\\nonumber\cdot
e^{-(m_{leptons}^2 s+\frac{\beta^2n^2}{4s})}(xs~Coth(xs)-1);
\end{eqnarray}

- quarks
\begin{eqnarray}
v^\prime_{quarks}=-\frac{1}{4 \pi^2}\sum_{f=1}^6\sum_{n=1}^{\infty}
(-1)^n \int_0^\infty
\frac{ds}{s^3}e^{-(m_{f}^2 s+\frac{\beta^2 n^2}{4s})}\\\nonumber
\cdot(q_fxs~Coth(q_fxs)\cdot ys~Coth(ys)-1);
\end{eqnarray}

- $W$-bosons (see \cite{EW})
\begin{eqnarray}
v^\prime_W=-\frac{x}{8\pi^2}\sum_{n=1}^\infty
\int_0^\infty\frac{ds}{s^2}e^{-(m_{W}^2 s+\frac{\beta^2
n^2}{4s})}\\\nonumber
\cdot\left[\frac{3}{Sinh(xs)}+4Sinh(xs)\right];
\end{eqnarray}

- gluons (see \cite{Once})
\begin{eqnarray}
\label{gluons}
v^\prime_{gluons}=-\frac{y}{4\pi^2}\sum_{n=1}^\infty
\int_0^\infty\frac{ds}{s^2}e^{-(m_{gluons}^2 s+\frac{\beta^2 n^2}{4s})}
\\\nonumber
\cdot\left[\frac{1}{Sinh(ys)}+
2Sinh(ys)\right].
\end{eqnarray}

Here, $m_{leptons}$, $m_{f}$, $m_{W}$ and $m_{gluons}$ are
the temperature masses of leptons, quarks, W-bosons and
gluons, respectively; $q_f=\left(\frac23,-\frac13,
-\frac13,\frac23,-\frac13,\frac23\right)$ - the charges of quarks.

Since we investigate the dynamics of high-temperature effects
connected with the presence of external fields, we used only the
leading in temperature terms of the Debye masses of the particles
(\cite{Once}, \cite{EW}).

The temperature masses of leptons and quarks are
\begin{eqnarray}
\begin{array}{cc}
m_{leptons}^2=\left(\frac{e}{\beta}\right)^2, &
m_{f}^2=\left(\frac{e}{\beta}\right)^2.
\end{array}
\end{eqnarray}

As it is known \cite{Once}, the transversal components of the
charged gluons and $W$-bosons have no temperature masses of order
$\sim g_s{\bf H}_3$ and $\sim gH$. Only the longitudinal
components have the Debye masses, but they are $H$- and ${\bf
H}_3$-independent, therefore, they can be omitted in our
consideration. Instead, the transversal component masses, which
depend on the Landau level number, must be used. So, the
transversal temperature masses of $W-$bosons and charged gluons
\begin{eqnarray}
\begin{array}{cc}
m_{W}^2= 15 \alpha_{e.w.}\frac{h^{1/2}}{ \beta}, &
m_{gluons}^2=15 \alpha_{s}\frac{h^{1/2}}{ \beta}
\end{array}
\end{eqnarray}
are to be substituted.
Here, $\alpha_{e.w.}$ and $\alpha_s$ are the electroweak and
the strong interaction couplings, correspondingly.

In the approximation adopted in the present
investigation we take as the masses the ground state energies of
the transversal modes \cite{SkSt}.

In the one-loop order the neutral gluon contribution is trivial
${\bf H}_3-$independent constant which can be omitted. However,
these fields are long-range states and they do give
${\bf H}_3-$dependent EP through the correlation corrections
depending on the temperature and field. We included only the
longitudinal neutral modes because their Debye's masses
$\Pi^0(y,\beta)$ are nonzero. The corresponding EP is
\cite{Once}
\begin{eqnarray}
\label{ring}
v_{ring}=\frac{1}{24\beta^2}\Pi^0(y,\beta)-\frac{1}{12\pi\beta}
\left(\Pi^0(y,\beta)\right)^{3/2}\\\nonumber
+\frac{\left(\Pi^0(y,\beta)\right)^2}{32\pi^2}
\left[log\left(\frac{4\pi}{\beta(\Pi^0(y,\beta))^{1/2}}\right)+\frac{3}{4}-\gamma\right],
\end{eqnarray}
$\gamma$ is Euler's constant,
$\Pi^0(y,\beta)=\Pi^0_{00}(k=0,y,\beta)$ is the zero-zero
component of the neutral gluon field polarization operator
calculated in the external field at finite temperature and taken
at zero momentum \cite{Once}
\begin{eqnarray}
\Pi^0(y,\beta)=\frac{2 g^2}{3
\beta^2}-\frac{y^{1/2}}{\pi\beta}-\frac{y}{4\pi^2}.
\end{eqnarray}
Equations (\ref{vprime})-(\ref{gluons}), (\ref{ring}) will be used
in numeric calculations.

\section{Generation of magnetic and chromomagnetic fields}
In order to find the strengths of generated magnetic and
chromomagnetic fields we have to find the minima of the EP in the
presence both of them. First of all we will find the strengths
$x$ and $y$ of fields, when the quark contribution is divided in
two parts
\begin{eqnarray}
v^\prime_{quarks}(x,\beta)=v^\prime_{quarks}\vert_{y\to0}\\\nonumber
=-\frac{1}{4 \pi^2}\sum_{f=1}^6\sum_{n=1}^{\infty} (-1)^n \int_0^\infty
\frac{ds}{s^3}e^{-(m_f^2 s+\frac{\beta^2 n^2}{4s})}\\\nonumber
\cdot(q_fxs~Coth(q_fxs)-1)
\end{eqnarray}
and
\begin{eqnarray}
v^\prime_{quarks}(y,\beta)=v^\prime_{quarks}\vert_{x\to0}\\\nonumber
=-\frac{1}{4 \pi^2}\sum_{f=1}^6\sum_{n=1}^{\infty} (-1)^n \int_0^\infty
\frac{ds}{s^3}e^{-(m_f^2 s+\frac{\beta^2 n^2}{4s})}\\\nonumber
\cdot(ys~Coth(ys)-1),
\end{eqnarray}
where ~ $v^\prime_{quarks}(x,\beta)$ is that one in the magnetic field,
$v^\prime_{quarks}(y,\beta)$ - in the presence of the chromomagnetic
field.

Let us rewrite the $v^\prime$ in (\ref{vprime}) as follows
\begin{eqnarray}
v^\prime(\bar{x},\bar{y})=v_1(\bar{x})+v_2(\bar{y})+v_3(\bar{x},\bar{y}),
\end{eqnarray}
where $\bar{x}=x+\delta x$, $\bar{y}=y+\delta y$, $\delta x$ and
$\delta y$ are the field corrections connected with the effect of
fields interfusion in the quark sector.

Since the mixing of fields due to a quark loop is weak (that will
be justified by numeric calculations) we can assume that
$\delta x<<1$ and $\delta y<<1$, and write
\begin{eqnarray}
\cases{
v_1(\bar{x})=v_1(x+\delta x)=v_1(x)+\frac{\partial v_1(x)}{\partial x}\delta
x,\cr
v_2(\bar{y})=v_2(y+\delta y)=v_2(y)+\frac{\partial v_2(y)}{\partial y}\delta
y,\cr
v_3(\bar{x},\bar{y})=v_3(x+\delta x,y+\delta y)=v_3(x,y).
}
\end{eqnarray}
After simple transformations we can find the $\delta x$ and $\delta y$
\begin{eqnarray}
\cases{
\delta x=\frac{\frac{\partial v_3(x,0)}{\partial x}-
\frac{\partial v_3(x,y)}{\partial x}}
{
\frac{\partial^2 v_1(x)}{\partial x^2}},\cr
\partial y=\frac{\frac{\partial v_3(0,y)}{\partial y}-
\frac{\partial v_3(x,y)}{\partial y}}{
\frac{\partial^2 v_2(y)}{\partial y^2}}.
}
\end{eqnarray}

Hence we may obtain $\bar{x}=x+\delta x$ and $\bar{y}=y+\delta y$.

The results on the field strengths determined by numeric investigation
of the total EP are summarized in Tables 1, 2.

In the first column of Tables 1 and 2 we show the inverse
temperature. In the second one the strength of magnetic and
chromomagnetic fields are adduced in the case of
quark EP, which describes each field separately. Next column gives
the field corrections in the case of total quark EP. The fourth
column presents the relative value of corrections. And the last
column gives the resulting strength of magnetic and chromomagnetic
fields, correspondingly.

\begin{figure}
\begin{center}
\resizebox{0.45\textwidth}{!}{\includegraphics{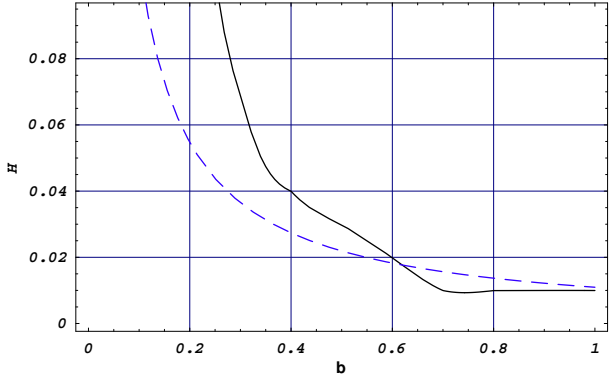}}
\caption{The dependence of the strengths of generated magnetic
field ($H$) on inverse temperature ($b$). The dashed line is the
theoretical position in the case of single magnetic field and the
solid is calculated one in presence of both fields.}
\label{fig:1}
\end{center}
\end{figure}

\begin{center}
\begin{tabular}{|c|c|c|c|c|}
\hline
$\beta$ & $x$ & $\delta x$ & $\delta x/x$, \% & $\bar{x}$ \\\hline\hline
0.1 & 0.7 & $0.0000165$ & $0.002$ & $0.7000165$\\
0.2 & 0.2 & $0.000745$ & $0.373$ & $0.200745$\\
0.3 & 0.07 & $-0.0000549$ & $-0.079$ & $0.0699451$\\
0.4 & 0.04 & $-0.0000358$ & $-0.090$ & $0.0399642$\\
0.5 & 0.03 & $-0.0000467$ & $-0.156$ & $0.0299533$\\
0.6 & 0.02 & $-0.0000492$ & $-0.246$ & $0.0199508$\\
0.7 & 0.01 & $-0.0000380$ & $-0.380$ & $0.0099620$\\
0.8 & 0.01 & $-0.0000619$ & $-0.619$ & $0.0099381$\\
0.9 & 0.01 & $-0.0000241$ & $-0.241$ & $0.0099759$\\
1.0 & 0.01 & $-0.0000357$ & $-0.357$ & $0.0099643$\\\hline
\end{tabular}
\end{center}

\noindent{{\bf Table 1.} {\it The strength of generated magnetic
field.}}
\vskip 1 cm

As it is seen, the increase of inverse temperature leads to
decreasing the strengths of generated fields. This dependence is
well accorded with the picture of  the universe cooling.

\begin{figure}
\begin{center}
\resizebox{0.45\textwidth}{!}{\includegraphics{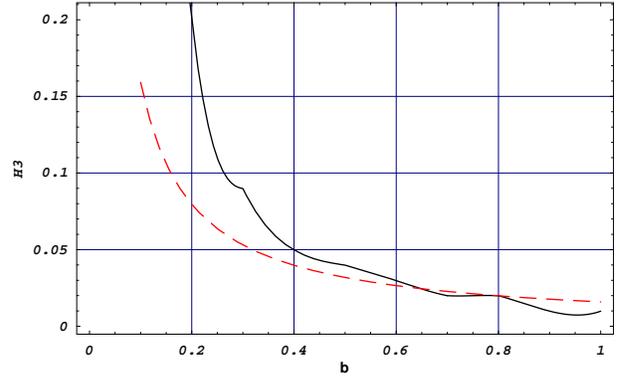}}
\caption{The dependence of the strengths of generated
chromomagnetic field ($H3$) on inverse temperature ($b$). The
dashed line is the theoretical position in the case of single
chromomagnetic field and the solid one is calculated  in the
presence of both fields.}
\label{fig:2}
\end{center}
\end{figure}

\vskip 1 cm
\begin{center}
\begin{tabular}{|c|c|c|c|c|}
\hline
$\beta$ & $y$ & $\delta y$ & $\delta y/y$, \% & $\bar{y}$ \\\hline\hline
0.1 & 0.8 & $0.000301$ & $0.038$ & $0.800301$ \\
0.2 & 0.2 & $-0.000239$ & $-0.119$ & $0.199761$ \\
0.3 & 0.09 & $-0.0000988$ & $-0.110$ & $0.0899012$ \\
0.4 & 0.05 & $-0.0000884$ & $-0.177$ & $0.0499116$ \\
0.5 & 0.04 & $-0.000112$ & $-0.280$ & $0.039888$ \\
0.6 & 0.03 & $-0.0000982$ & $-0.327$ & $0.0299018$ \\
0.7 & 0.02 & $-0.0000442$ & $-0.221$ & $0.0199558$ \\
0.8 & 0.02 & $-0.0000733$ & $-0.367$ & $0.0199267$ \\
0.9 & 0.01 & $-0.000117$ & $-1.166$ & $0.009883$ \\
1.0 & 0.01 & $-0.000175$ & $-1.749$ & $0.009825$ \\\hline
\end{tabular}
\end{center}

\noindent{{\bf Table 2.} {\it The strengths of generated
chromomagnetic field.}}
\vskip 1 cm

From the above analysis it follows that at high temperatures the
value of the each type magnetic field is increased when other one is
taken into account. With temperature decreasing this effect
becomes less pronounced and disappears at comparably low
temperatures $\beta \sim 1$.

\section{Discussion}

Let us discuss the results obtained. As it was elaborated in the
approximation to the EP including the one-loop and the daisy
diagrams, in the SM at high temperatures both the magnetic and
chromomagnetic fields have to be generated. These states are
stable, as it follows from the absence of the  imaginary terms in
the EP minima.

If the quark loops are discarded, both of the fields can be
generated in the system, separately. All these states are stable,
due to magnetic mass $\sim g^2 (gH)^{1/2}T$ of transversal gauge
field modes. Here it worth to mention that the one-loop
transversal gauge field mass is of order $ \sim g^4 T^2$ as
nonperturbative calculations predict. This estimate is because the
magnetic field strength of the  spontaneously generated fields is
of order $(gH)^{1/2} \sim g^2 T$ \cite{Sta}, \cite{Once}. The
possibility to calculate the magnetic mass in perturbation theory
is due to the approach when an external field is taken into
consideration exactly when the polarization operator of gauge field
is calculated \cite{SkSt}. If one accounts for the magnetic field
perturbatively, zero value will be obtained \cite{Pers}.

As it is seen from the Figures 1,2, presenting the results of
numeric computations within the exact EP, the strengths of
generated fields are increasing with the temperature rising. It is
also found that the curves obtained in high temperature expansion
of the EP \cite{Once} are in good agreement with our numeric
calculations.

 The ground state possessing the magnetic
and chromomagnetic fields makes advantage for existing of these
fields in the electroweak transition epoch. The state with the
fields is stable in the whole considered temperature interval. The
imaginary part in the EP exists for the external fields much
stronger then the strengths of the spontaneously generated ones.
 The interfusion of magnetic and chromomagnetic fields arisen from
the quark sector of the EP is weak. The change of the field
minima in inclusion of the fields mixing does not exceed $2$
per cents.

 During the cooling of the universe the strengths of generated
fields are decreasing  that is in an agreement with  what is
expected in cosmology.

One of the consequences on the results obtained is the presence of
strong chromomagnetic field in the early universe, in particular,
at the electroweak phase transition and, probably, till the
deconfinement temperature. Influence of this field on the
transitions may bring new insight on these problems. As our
estimate showed, the chromomagnetic field is as strong as
magnetic one. So, the role of strong interactions in the early
universe in the field presence needs more detailed investigations
as compared to what is usually assumed \cite{Dar}.

\section*{Acknowledgements}

One of the authors (VS) thanks the Abdus Salam International
Center for Theoretical Physics, Trieste, Italy, where the final
part of this work was done, for hospitality.

\end{document}